\documentclass[aps,prl,reprint,superscriptaddress,nobibnotes,nofootinbib]{revtex4-2}
\usepackage[T1]{fontenc}
\usepackage{newtxtext,newtxmath}

\usepackage{graphicx}
\usepackage{dcolumn}
\usepackage{float}
\usepackage{placeins}
\usepackage{multirow}
\usepackage[normalem]{ulem}
\usepackage{color}
\usepackage{amsmath}
\usepackage{hyperref}
\usepackage{cleveref}
\newcommand{\dd}{\mathrm{d}}
\usepackage{bm}

\begin{document}

\title{Accurate recovery of the two linewidths hidden in laser beatnote statistics}
\author{Jingming Chen}
\thanks{These authors contributed equally to this work.}
\affiliation{School of Electronics, Peking University, Beijing, 100871, China}
\author{Yuanchen Qi}
\thanks{These authors contributed equally to this work.}
\affiliation{School of Electronics, Peking University, Beijing, 100871, China}
\author{Yuzheng Pang}
\affiliation{School of Electronics, Peking University, Beijing, 100871, China}
\author{Jie Miao}
\affiliation{School of Electronics, Peking University, Beijing, 100871, China}
\author{Congyu Wang}
\affiliation{\mbox{State Key Laboratory of Precision Spectroscopy, East China Normal University, Shanghai, 200062, China}}
\author{Yuan Yao}
\affiliation{\mbox{State Key Laboratory of Precision Spectroscopy, East China Normal University, Shanghai, 200062, China}}
\author{Zhi-Ang Chen}
\affiliation{International Center for Quantum Materials, School of Physics, Peking University, Beijing 100871, China}
\author{Run-Qi Lei}
\affiliation{International Center for Quantum Materials, School of Physics, Peking University, Beijing 100871, China}
\author{Zheyi Ge}
\affiliation{School of Electronics, Peking University, Beijing, 100871, China}
\affiliation{National Key Laboratory of Advanced Micro and Nano Manufacture Technology, Beijing, 100871, China}
\author{Yanyi Jiang}
\affiliation{\mbox{State Key Laboratory of Precision Spectroscopy, East China Normal University, Shanghai, 200062, China}}
\affiliation{Hefei National Laboratory, Hefei, 230088, China}
\author{Xibo Zhang}
\affiliation{International Center for Quantum Materials, School of Physics, Peking University, Beijing 100871, China}
\affiliation{Hefei National Laboratory, Hefei, 230088, China}
\author{Xiaopeng Xie}
\affiliation{School of Electronics, Peking University, Beijing, 100871, China}
\affiliation{State Key Laboratory of Photonics and Communications, Peking University, Beijing 100871, China}
\author{Jianjun Wu}
\affiliation{School of Electronics, Peking University, Beijing, 100871, China}
\affiliation{State Key Laboratory of Photonics and Communications, Peking University, Beijing 100871, China}
\author{Duo Pan}
\email{panduo@pku.edu.cn}
\affiliation{School of Electronics, Peking University, Beijing, 100871, China}
\affiliation{National Key Laboratory of Advanced Micro and Nano Manufacture Technology, Beijing, 100871, China}
\author{Jingbiao Chen}
\affiliation{School of Electronics, Peking University, Beijing, 100871, China}
\affiliation{National Key Laboratory of Advanced Micro and Nano Manufacture Technology, Beijing, 100871, China}
\affiliation{Hefei National Laboratory, Hefei, 230088, China}
\begin{abstract}
Photon from ultrastable lasers can remain coherent length over distances approaching the Earth–Sun separation, enabling optical clocks projected to lose less than one second over the age of the Universe. Yet characterizing these photons poses an identifiability problem: a two-laser heterodyne measurement produces a single beatnote linewidth that conflates photons contributions from both lasers. For more than half a century, the standard solution has been the three-cornered-hat (TCH) method, which requires three independent ultrastable lasers. Here we derive the mathematical form and elucidate the physical origin of asymmetric function of beatnote-linewidth distributions arising from finite photon wave trains, a long-observed feature not captured by canonical Gaussian or Lorentzian statistics. This finding accurately recovers two linewidths hidden in laser beatnote statistics without a third laser. The framework consistently captures the observed coherent length and time statistics of photons, while comparison with TCH measurements confirms the quantitative validity of the extracted individual linewidths across five independent ultrastable-laser datasets spanning nearly two orders of magnitude. Most notably, the method resolves the 7.8-mHz linewidth of a cryogenic silicon-cavity laser beating with a broader one. This transformative function of beatnote-linewidth distributions, together with the resulting method, could fundamentally advance optical clocks and precision metrology.
\end{abstract}
\date{\today}
\maketitle
\sloppy
\textit{Introduction\textemdash}Ultra-narrow-linewidth lasers constitute a fundamental enabling backbone for state-of-the-art precision metrology, underpinning optical atomic clocks~\cite{ludlow2015optical,king2022Optical,mcgrew2018Atomic,jiang2011making,nicholson2012comparison,zhang2026liquid,lin202187sr,collaboration2021frequency,oelker2019demonstration,le2013experimental}, gravitational-wave detection~\cite{tse2019quantum,kolkowitz2016gravitational}, high-resolution spectroscopy~\cite{van2011frequency} and ultra-low–phase-noise microwave generation~\cite{fortier2011generation,he2025highly}. Continuous advances in laser-linewidth narrowing techniques have pushed achievable linewidths into the millihertz regime~\cite{matei20171,chen2025laser,lee2025frequency,bohnet2012steady,norcia2018frequency,gundavarapu2019sub,xiao2026Continuouswave,Schioppo2026EFTF,robinson2019crystalline}, enabled by cryogenic silicon cavities~\cite{jiang2011making,zhu2024ultrastable,zhu2025transportable}, spectral-hole burning~\cite{thorpe2011frequency}, and active optical clock superradiant lasing~\cite{yu2007optical,chen2009active,meiser2009prospects,dubey2025modeling}. As laser coherence reaches unprecedented levels, accurate linewidth characterization has become increasingly critical and challenging.
\par
Accurate linewidth characterization most commonly relies on the three-cornered-hat (TCH) method~\cite{gray1974method}, which extracts individual linewidths from pairwise beatnote measurements among three comparable lasers. Despite its widespread use, the TCH method intrinsically requires three lasers with closely matched performance~\cite{matei20171,chen2025laser,lee2025frequency,zhang2017ultrastable}, demanding substantial technical effort, financial resources, and time commitment, thereby limiting its applicability in typical laboratory settings. A simpler and more widely adopted alternative is the two-laser heterodyne method, which assumes equal contributions from the two lasers to the beatnote linewidth~\cite{wu20160,liang2015ultralow,corato2023widely,bian20161,hao2024stability} and consequently cannot accurately extract their individual linewidths.
\par
\begin{figure*}[!htb]
    \centering
    \includegraphics[width=0.90\textwidth,height=0.21\textwidth]{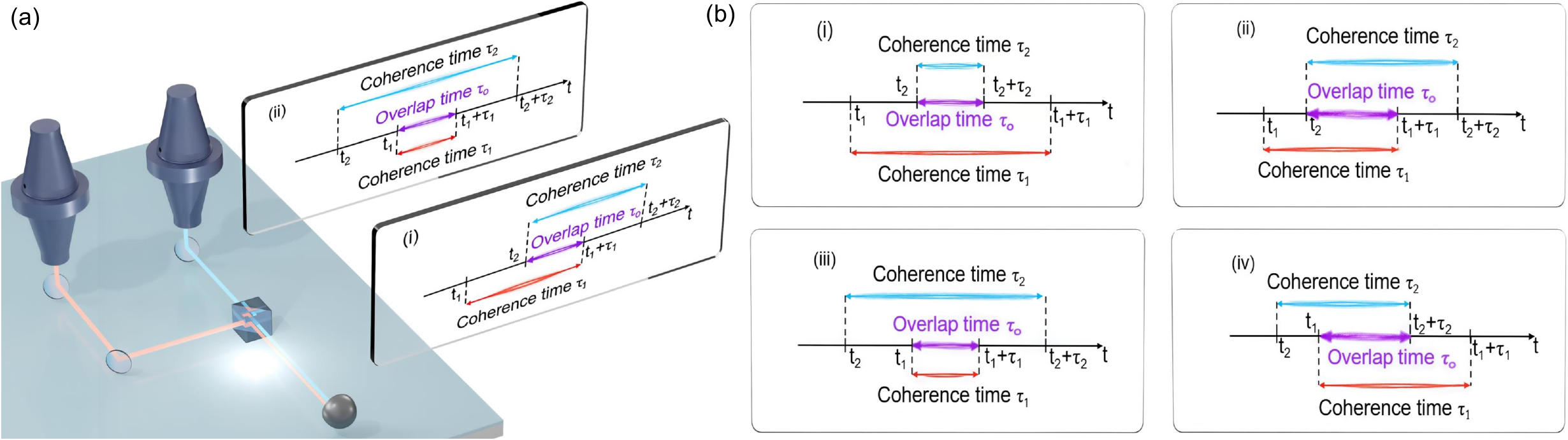}
    \caption{Physical model of optical heterodyne detection. (a) Schematic diagram of the beating between two lasers. Insets (i) and (ii) illustrate two typical conditions, where the coherence times of the two lasers are significantly different or approximately equal. (b) Possible temporal configurations of the coherence wave trains of the two lasers. (i) and (ii) correspond to the scenario where the coherence of laser 1 starts earlier than that of laser 2, while (iii) and (iv) correspond to the opposite scenario. Based on the characteristics of the overlap time, cases (i) and (iii) exhibit a situation where the coherence time of one laser fully encompasses that of the other, which we refer to as the “coverage” case. In contrast, cases (ii) and (iv) always exhibit a situation where the coherence times of the two lasers only partially overlap, which we refer to as the “partial overlap” case.}
    \label{1fig}
\end{figure*}
A fair and reliable characterization of laser linewidths based on heterodyne beat measurements requires accounting for the statistical nature of the beatnote process. L. S. Ma first emphasized this aspect and proposed characterizing laser linewidths using statistical ensembles of repeated beatnote measurements, with the most probable linewidth serving as the estimator~\cite{jiang2010nd}. Over the past sixteen years, this approach has been widely adopted, with numerous studies employing beatnote linewidth distributions (BLDs) to characterize laser coherence properties~\cite{wu20160,zhang2017ultrastable,matei20171,wang2020integrated,bian20161,hao2024stability}. However, experimentally measured BLDs exhibit pronounced and anomalous asymmetries that differ markedly from classical symmetric profiles such as Gaussian or Lorentzian functions, which fail to capture both the observed statistics and the underlying physics~\cite{hao2024stability,wu20160}. This limited applicability of conventional lineshape descriptions bears some resemblance to early studies of blackbody radiation, where neither Wien’s law nor the Rayleigh–Jeans law could fully and accurately describe the entire lineshape before Planck’s law provided a universal description of blackbody radiation. A physically grounded statistical description of the BLD is therefore required.
\par
In this Letter, we introduce a beatnote-linewidth distribution function, termed the \(\mathcal{M}a\) function in recognition of L. S. Ma’s seminal contributions~\cite{jiang2010nd}, that enables simultaneous extraction of the individual linewidths of two lasers by fitting experimentally measured BLDs. First, we show that the asymmetry of BLDs originates from the partial overlap of photon-coherence wave trains. This physical picture establishes a direct mapping between BLDs and the intrinsic linewidths of the individual lasers. Second, the \(\mathcal{M}a\) function accurately fits multiple experimental BLDs, including those measured using state-of-the-art millihertz-linewidth lasers~\cite{wu20160,chen2025laser,zhang2017ultrastable,matei20171,hao2024stability}, and yields individual linewidths consistent with those obtained using the conventional TCH method. Finally, \(\mathcal{M}a\)-function fitting resolves the narrower linewidth of a cryogenic-cavity-stabilized laser from its BLD with a broader-linewidth room-temperature-cavity reference, even at a substantial linewidth ratio between the two lasers. This work establishes a physically grounded and experimentally accessible method for laser linewidth characterization without the complexity and high cost of the conventional TCH method.
\par
\textit{Theoretical model\textemdash}To model the BLD, we adopt a time-domain framework commonly used in laser noise analysis and formulate a physical description based on the laser coherence time. In an ideal system, the coherence time \(\tau\) is a well-defined parameter that characterizes the exponential decay of the field correlation function. In practical systems, however, frequency noise spanning multiple time scales leads to stochastic phase fluctuations, rendering the coherence time a statistical variable. We therefore introduce a statistical distribution \(f(\tau)\) to describe its statistics.
\par
When two lasers are heterodyned, the coherence properties of the beatnote signal are governed by the evolution of their relative optical phase. The observed linewidth depends on the effective overlap of coherence times of the two lasers, as illustrated schematically in Fig.~\ref{1fig}(a), where each laser is represented as a finite coherence wave train. Insets (i) and (ii) illustrate two representative beating regimes. When the two lasers have comparable linewidths, their coherence times are of similar magnitude. During the beating process, the overlap time of their coherence wave trains is then typically shorter than either individual coherence time. Because linewidth is inversely related to coherence time, the resulting beatnote linewidth is therefore likely to exceed the linewidths of both lasers. By contrast, when the linewidths differ substantially, the shorter coherence time of the broader-linewidth laser limits the overlap time, causing the beatnote linewidth to approach that of the broader-linewidth laser.
\begin{figure}
    \centering
    \includegraphics[width=6.8cm,height=6.2cm]{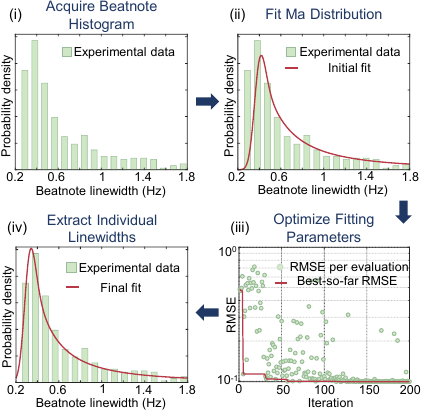}
    \caption{Workflow for extracting individual laser linewidths from a BLD. Beatnote linewidth samples are used to construct the experimental BLD. The \(\mathcal{M}a\) function is then fitted to the measured distribution. The mean linewidths \(\bar{\gamma}_i\) and standard deviations \(\sigma_i\) are optimized using a Bayesian optimization algorithm that minimizes the RMSE. The optimal parameters provide the individual linewidths and standard deviations of the two lasers.}
    \label{workflow}
\end{figure} 
\par
We denote the overlap time by \(\tau_o\) and define the corresponding beatnote linewidth as its inverse, i.e., \(\gamma_\text{beat}=1/\tau_o\), following the Fourier relation between coherence time and linewidth. The problem of determining the statistical distribution of \(\gamma_\text{beat}\) can thus be reformulated as finding the distribution of the overlap time \(\tau_o\). Each laser is modeled as a coherence wave train of duration \(\tau_1\) and \(\tau_2\), starting at times \(t_1\) and \(t_2\), respectively. The overlap of the two wave trains on the time axis defines the overlap time \(\tau_o\) during which the two optical fields remain phase correlated in the heterodyne process, as illustrated in Fig.~\ref{1fig}(b). Specifically, the first wave train occupies the interval \((t_1, t_1+\tau_1)\), while the second occupies \((t_2, t_2+\tau_2)\). A nonzero overlap occurs when the two intervals intersect, i.e., \((t_1-\tau_2)<t_2<(t_1+\tau_1)\). Without loss of generality, we set \(t_1=0\), reducing the overlap condition to \(-\tau_2<t_2<\tau_1\). Under this condition, the relative temporal arrangements of the two wave trains can be classified into four distinct cases, as shown in Fig.~\ref{1fig}(b). For each case, the corresponding overlap time \(\tau_o\) is given by 
\begin{align}
    \tau_o=\begin{cases}
        \tau_2; \ &\mathrm{case}\ (\mathrm{i}):\ 0<t_2<t_2+\tau_2<\tau_1;\\
        \tau_1-t_2; \ &\mathrm{case}\ (\mathrm{ii}):\ 0<t_2<\tau_1<t_2+\tau_2;\\
        \tau_1; \ &\mathrm{case}\ (\mathrm{iii}):\ t_2<0<\tau_1<t_2+\tau_2;\\
        \tau_2+t_2; \ &\mathrm{case}\ (\mathrm{iv}):\ t_2<0<t_2+\tau_2<\tau_1.\\
    \end{cases} 
    \label{1}
\end{align}
When one wave train fully contains the other [cases (i) and (iii)], this situation is referred to as ``coverage''; otherwise [cases (ii) and (iv)], it corresponds to ``partial overlap''. The statistical distribution of the overlap time \(\tau_o\) is then obtained by evaluating the cumulative probability and taking its derivative for each case~\cite{durrett2019probability}. Summing the contributions from all four cases yields the total statistical distribution of the overlap time, denoted by \(f_o(\tau_o)\), which can be written as
\begin{equation}
    \begin{alignedat}{3}
        f_{o}(\tau_o)=&f_2(\tau_o)\int_{\tau_o}^{+\infty}\frac{\tau_1-\tau_o}{\tau_1+\tau_o}f_1(\tau_1)\dd \tau_1 & \ (\mathrm{i});\\
        +&\int_{\tau_o}^{+\infty}\left(\int_{\tau_o}^{+\infty}\frac{1}{\tau_1+\tau_2}f_1(\tau_1)f_2(\tau_2)\dd\tau_1\right)\dd\tau_2 &\ (\mathrm{ii});\\
        +&f_1(\tau_o)\int_{\tau_o}^{+\infty}\frac{\tau_2-\tau_o}{\tau_2+\tau_o}f_2(\tau_2)\dd \tau_2 &\ (\mathrm{iii});\\
         +&\int_{\tau_o}^{+\infty}\left(\int_{\tau_o}^{+\infty}\frac{1}{\tau_1+\tau_2}f_1(\tau_1)f_2(\tau_2)\dd\tau_1\right)\dd\tau_2&(\mathrm{iv}).
          \label{2}
    \end{alignedat}
\end{equation}
Here, \(f_i(\tau_i)\) denotes the statistical distribution of the coherence time of the \(i\)-th laser, and each term corresponds directly to one of the overlap-time expressions listed in Eq.(\ref{1}). To facilitate direct extraction of the individual laser linewidths from a BLD, we further introduce a dedicated linewidth distribution (LD) \(g(\gamma_i)\) for the \(i\)-th laser, which reflects the statistical variation of its measured linewidth over multiple realizations. For the ultrastable lasers studied here, we assume a Gaussian distribution for the linewidth of each laser,
\begin{align}
    g(\gamma_i)=\dfrac{1}{\sqrt{2\pi}\sigma_i}\exp(-\dfrac{(\gamma_i-\bar\gamma_i)^2}{2\sigma_i^2}).\label{gau}
\end{align}
Here, \(\bar\gamma_i\) is the mean linewidth of the \(i\)-th laser and the standard deviation \(\sigma_i\) characterizes the linewidth fluctuations induced by environmental perturbations and technical noise. Owing to the Fourier relation between coherence time and linewidth, \(\gamma\simeq1/\tau\), the coherence-time distribution \(f(\tau)\) is then obtained from the LD \(g(\gamma)\) as \(f(\tau)=g(1/\tau)/\tau^2\). Transforming the overlap-time distribution \(f_o(\tau_o)\) back to the linewidth domain yields the statistical distribution function of the beatnote linewidth, referred to here as the \(\mathcal{M}a\) function, i.e.,
\begin{align}
\mathcal{M}_a(\gamma_{\text{beat}})=\frac{f_{o}(1/\gamma_{\text{beat}})}{(\gamma_{\text{beat}})^2}.
    \label{3}
\end{align}
Although the \(\mathcal{M}a\) function is expressed here in integral form, future analysis under specific limiting conditions may yield analytical solutions and provide deeper physical insight.
\par
\begin{figure}
    \centering
    \includegraphics[width=7.4cm,height=12.6cm]{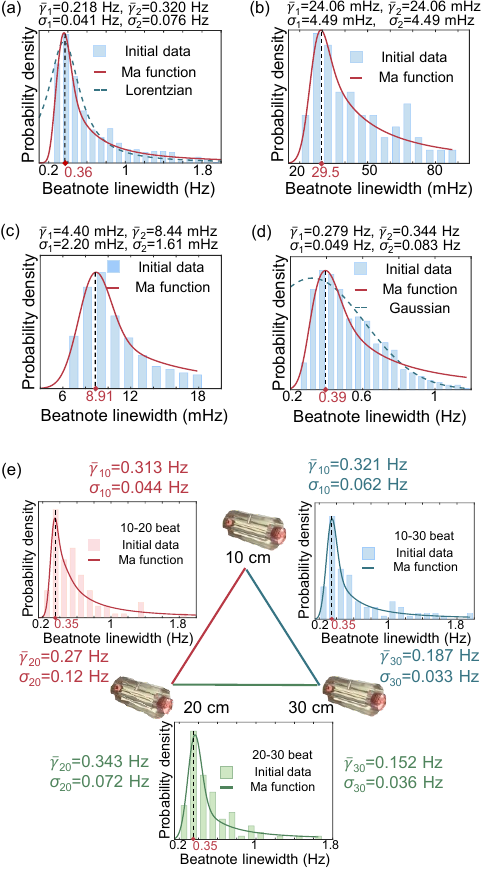}
    \caption{\label{fig10new}\(\mathcal{M}a\)-function fits to experimental BLDs. The LD of each laser is assumed to follow a Gaussian distribution, with mean values \(\bar{\gamma}_{1,2}\) and standard deviations \(\sigma_{1,2}\). Panels (a)–(d) show four independent experimental BLDs. (a) Two 1557 nm diode lasers locked to two 10 cm room-temperature Fabry–Pérot cavities~\cite{wu20160}:  \(\bar{\gamma}_1=0.218~{\rm Hz}\), and \(\bar{\gamma}_2=0.320~{\rm Hz}\) (RMSE=0.0954). (b) Comb-mediated beatnote at 771 nm between a silicon-cavity-stabilized laser and a 40 cm ULE-cavity-stabilized laser~\cite{zhang2017ultrastable}: \(\bar{\gamma}_1=24.06~{\rm mHz}\), and \(\bar{\gamma}_2=24.06~{\rm mHz}\) (RMSE=0.0023), equivalent to 12.03 mHz at 1542 nm and consistent with the reported 12.5 mHz most probable linewidth and 17 mHz median linewidth. (c) Two cryogenic silicon-cavity-stabilized lasers operating at 124 K~\cite{matei20171}: \(\bar{\gamma}_1=4.40~{\rm mHz}\), and \(\bar{\gamma}_2=8.44~{\rm mHz}\) (RMSE=0.0033). (d) Beatnote between two 729 nm Ti:sapphire laser paths locked to two 30 cm ULE cavities~\cite{hao2024stability}: \(\bar{\gamma}_1=0.279~{\rm Hz}\), and \(\bar{\gamma}_2=0.344~{\rm Hz}\) (RMSE=0.1735). (e) TCH-type validation experiment using two 1064 nm lasers stabilized to 10 cm and 20 cm ULE cavities and one 578 nm laser stabilized to a 30 cm ULE cavity. The 10–20, 10–30, and 20–30 beat BLDs are shown in red, blue, and green, respectively, together with their \(\mathcal{M}a\)-function fits.}
\end{figure}
\par
Building on the theoretical framework developed above, the BLD can be obtained experimentally through repeated heterodyne measurements between two lasers. By fitting the measured BLD with the \(\mathcal{M}a\) function, the individual linewidths of the two lasers can be directly extracted. To enable efficient implementation, we have developed a lightweight software tool~\cite{2026softurl} for automatic fitting, with the workflow summarized in Fig.~(\ref{workflow}). Specifically, (i) a sufficiently large number of beatnote linewidth samples is acquired to construct a statistically reliable BLD, as suggested by Ma~\cite{jiang2010nd}. (ii) The data are imported into the software, where the LD of each laser is modeled by the Gaussian distribution defined in Eq.(\ref{gau}). The mean linewidths \(\bar\gamma_1\) and \(\bar\gamma_2\), as well as the corresponding standard deviations \(\sigma_1\) and \(\sigma_2\), are treated as fitting parameters in the \(\mathcal{M}a\)-function-based analysis. (iii) An embedded Bayesian optimization algorithm iteratively updates these four parameters by minimizing the root-mean-square error (RMSE) between the \(\mathcal{M}a\) function and the experimental BLD~\cite{jones1998efficient}. (iv) From the optimal fit, the software returns the mean linewidths \(\bar\gamma_i\) of the two lasers together with their standard deviations \(\sigma_i\). By construction, the statistical analysis itself does not intrinsically assign a given linewidth to a specific laser. This correspondence must be identified by the user based on additional experimental knowledge, such as prior knowledge of which laser exhibits the narrower linewidth.
\par
To validate the proposed model, we apply the \(\mathcal{M}a\) function to four experimental BLDs obtained by three research groups, including state-of-the-art millihertz-linewidth systems~\cite{wu20160,zhang2017ultrastable,matei20171,hao2024stability}. The results are shown in Fig.~\ref{fig10new}(a-d), where each BLD is presented as a histogram. The corresponding \(\mathcal{M}a\)-function fits (solid curves) agree closely with the experimental data over the full distribution range. For comparison, the conventional function fits reported in the original references---Lorentzian in Fig.~\ref{fig10new}(a) and Gaussian in Fig.~\ref{fig10new}(d)---are shown as blue dashed curves. In both comparisons, the \(\mathcal{M}a\) function yields a smaller RMSE and more accurately captures the statistical structure of the BLDs, enabling reliable extraction of the individual laser linewidths. We further validate the \(\mathcal{M}a\)-function analysis using a TCH experiment involving two 1064 nm lasers stabilized to 10 cm and 20 cm ULE cavities and a 578 nm laser stabilized to a 30 cm ULE cavity, as shown in Fig.~\ref{fig10new}(e). The resulting pairwise BLDs exhibit nearly indistinguishable most probable linewidths of approximately 0.35 Hz, whereas conventional TCH analysis yields nealy identical individual linewidths of about 0.25 Hz. \(\mathcal{M}a\)-function fitting resolves their otherwise hidden linewidth ratios among the three lasers and distinguishes their relative performance. For each laser, the two pairwise beat measurements provide independent \(\mathcal{M}a\)-function estimates of its linewidth. Averaging these estimates gives linewidths of approximately 0.32, 0.31, and 0.17 Hz for the 10 cm, 20 cm, and 30 cm cavity lasers, respectively. The linewidth values extracted for the same laser from different beatnote pairs are generally consistent, with residual discrepancies likely arising from environmental perturbations, variations in locking conditions, and finite sampling. In practice, reliable \(\mathcal{M}a\)-function fitting requires sufficient sampling and an appropriate instrumental resolution bandwidth (RBW) to fully resolve the beatnote linewidth without introducing additional broadening. These results demonstrate that linewidth extraction using the \(\mathcal{M}a\) function is consistent with conventional TCH method while providing additional statistical information about the individual lasers.
\par
\begin{figure*}[!htb]
    \centering
    \includegraphics[width=14cm,height=8.4cm]{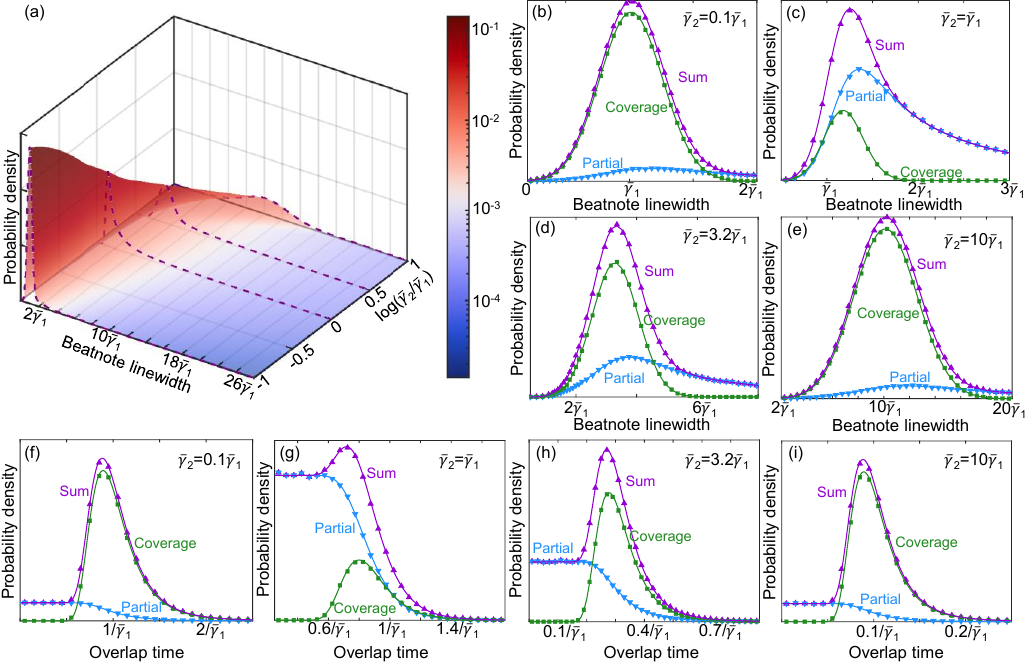}
    \caption{\label{fig10neweng} Evolution of the \(\mathcal{M}a\) function as the mean linewidth \(\bar{\gamma}_2\) of laser 2 increases while that of laser 1, \(\bar{\gamma}_1\), remains fixed. The LD of each laser is assumed to be Gaussian, with \(\sigma_i=\bar{\gamma}_i/4\). (a) Three-dimensional visualization of the \(\mathcal{M}a\) function as a function of the beatnote linewidth and the linewidth ratio \(\log_{10}(\bar{\gamma}_2/\bar{\gamma}_1)\)., (b)-(e) \(\mathcal{M}a\) functions for linewidth ratios \(\bar{\gamma}_2/\bar{\gamma}_1=\) 0.1, 1, $\sqrt{10}$, and 10, respectively. The total \(\mathcal{M}a\) function (purple triangles) is decomposed into contributions from partial overlap (blue downward triangles) and coverage (green squares). Solid curves of the corresponding colors represent the theoretical predictions. (f)-(i) Statistical distributions of the overlap time corresponding to the linewidth ratios in panels (b)-(e), respectively, illustrating the relative contributions of partial overlap and coverage to the asymmetry of the \(\mathcal{M}a\) function.}
\end{figure*}
\par
\textit{Numerical investigation\textemdash}The linewidth ratio between the two lasers in a heterodyne measurement plays a central role in shaping the \(\mathcal{M}a\) function. To quantify this effect, we fix the mean linewidth of the first laser, \(\bar\gamma_1\), and progressively increase that of the second laser, \(\bar\gamma_2\), as shown in Fig.~\ref{fig10neweng}(a), assuming Gaussian LDs with \(\sigma_i=0.25\bar\gamma_i\). When the two linewidths are comparable, the resulting \(\mathcal{M}a\) function exhibits pronounced asymmetry, fundamentally distinct from symmetric Gaussian or Lorentzian functions, as illustrated in Fig.~\ref{fig10neweng}(c). As the linewidth ratio increases, the contribution from partial overlap diminishes, and the beatnote statistics become increasingly dominated by coverage. Consequently, the \(\mathcal{M}a\) function gradually reflects the characteristics of the broader-linewidth laser, while still retaining sufficient asymmetry to allow reliable extraction of both linewidths, as shown for a tenfold linewidth ratio in Fig.~\ref{fig10neweng}(b) and Fig.~\ref{fig10neweng}(e). The persistence of this residual asymmetry suggests that \(\mathcal{M}a\)-function fitting may retain sensitivity to individual linewidths even as the linewidth mismatch approaches an order of magnitude. For sufficiently large linewidth ratios, the \(\mathcal{M}a\) function ultimately converges to the LD of the broader-linewidth laser alone. In this limit, the beatnote statistics no longer contain sufficient information to resolve the narrower linewidth, which defines the practical boundary of applicability for the present method.
\par
To further clarify the temporal origin of the beatnote-linewidth statistics, we examine the overlap-time distributions associated with the four linewidth-ratio cases in Fig.~\ref{fig10neweng}(b-e), as shown in Fig.~\ref{fig10neweng}(f-i). Using the mean linewidths \(\bar\gamma_1\), \(\bar \gamma_2\) and the standard deviations \(\sigma_1\), \(\sigma_2\) defined in the simulations, we perform Monte Carlo simulations with 10,000 randomly generated wave-train pairs. For each realization, the overlap time \(\tau_o\) is calculated and classified as either ``coverage'' or ``partial overlap'', allowing the respective contributions to be directly decomposed. The simulation results are shown as green squares (coverage) and blue downward triangles (partial overlap) in Fig.~\ref{fig10neweng}(f-i). The theoretical predictions from Eq.(\ref{2}), shown as solid curves, are in excellent agreement with the simulation data, further confirming the statistical consistency and reliability of the model.
\par
The decomposition of the overlap-time statistics reveals the distinct roles of the ``coverage'' and ``partial overlap'' processes. When the linewidths of the two lasers are comparable, the overlap-time distribution is dominated by ``partial overlap'', as illustrated in Fig.~\ref{fig10neweng}(g). As the linewidth ratio increases, the ``coverage'' contribution becomes dominant, and the overlap time is progressively governed by the shorter coherence time of the broader-linewidth laser, as shown in Fig.~\ref{fig10neweng}(h–i). Notably, the contribution of ``partial overlap'' is approximately uniform at small \(\tau_o\) and rapidly decays at larger overlap times, as indicated in Fig.~\ref{fig10neweng}(g). This characteristic temporal behavior provides the primary physical origin of the pronounced asymmetry observed in the BLDs. The near-uniform distribution at small \(\tau_o\) arises from events in which the two wave trains partially overlap only at their edges. In this regime, the overlap time \(\tau_o\) becomes independent of the coherence-time distribution and is instead determined solely by the starting-time distribution. Since the starting time \(t_2\) is uniformly distributed over \((-\tau_2,\tau_1)\), the overlap time \(\tau_o\) inherits this uniformity and yields a constant probability density of \(2\bar\gamma_1\bar\gamma_2/(\bar\gamma_1+\bar\gamma_2)\). As a result, the \(\mathcal{M}a\) function exhibits an inverse-square tail at large beatnote linewidths, scaling as \( 1/\gamma_{\text{beat}}^2\).
\par
\begin{figure}[!htb]
    \centering
    \includegraphics[width=6.4cm,height=9.5cm]{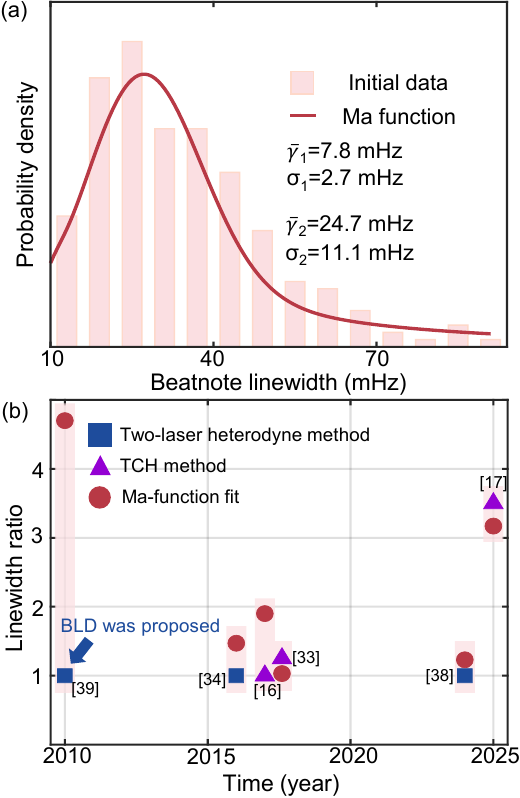}
    \caption{\label{fig5}(a) Experimental BLD obtained from a 1397 nm cryogenic silicon-cavity-stabilized laser and a room-temperature ULE-cavity-stabilized laser system, together with the \(\mathcal{M}a\)-function fit. The extracted linewidths are \(\bar{\gamma}_1=7.8\) mHz and \(\bar{\gamma}_2=24.7\) mHz. (b) Comparison of linewidth ratios obtained using the conventional two-laser heterodyne method, the TCH method, and \(\mathcal{M}a\)-function fitting. Marker shapes distinguish the analysis methods used to infer the linewidth ratios, while shaded regions identify datasets for which \(\mathcal{M}a\)-function fitting is directly compared with the corresponding originally reported analysis. For Ref.~\cite{zhang2017ultrastable}, the plotted value denotes the equivalent linewidth ratio of the two cavity-stabilized lasers at 771 nm.}
\end{figure}

The linewidth-ratio analysis above identifies a practical reference-limited regime in which the laser under test is appreciably narrower than the available independent reference, yet the BLD remains sensitive to its contribution even when the beatnote is dominated by the broader-linewidth laser. This regime is particularly relevant to cryogenic-cavity-stabilized lasers, whose full TCH characterization would require multiple independent millihertz-linewidth references~\cite{zhang2017ultrastable,matei20171,chen2025laser,Schioppo2026EFTF,robinson2019crystalline}. \(\mathcal{M}a\)-function fitting avoids this requirement by extracting the narrow-linewidth laser contribution encoded in the BLD. We therefore perform a new set of repeated heterodyne measurements between the 1397 nm cryogenic silicon-cavity-stabilized laser and the room-temperature ULE-cavity-stabilized reference reported in Ref.~\cite{chen2025laser}. The resulting BLD, comprising 226 data points, is then fitted with the \(\mathcal{M}a\) function, as shown in Fig.~\ref{fig5}(a). The fit yields individual linewidths of \(\bar{\gamma}_1(\sigma_1)=7.8\) (2.7)  mHz and \(\bar{\gamma}_2(\sigma_2)=24.7\) (11.1) mHz, corresponding to a linewidth ratio of 3.17. The fitting results are consistent with the previous TCH-based analysis of the same laser system, with \(\bar{\gamma}_1\) being fairly consistent with the reported 9.6 mHz for the 1397 nm cryogenic-silicon-cavity laser and the linewidth ratio agreeing with the value of 3.56 estimated from the data in Ref.~\cite{chen2025laser}. This result validates \(\mathcal{M}a\)-function fitting in a reference-limited heterodyne configuration, showing that the narrower linewidth can be recovered from an unequal-linewidth beat without requiring a metrologically equivalent TCH ensemble.
\par
Furthermore, Fig.~\ref{fig5}(b) compares different characterization methods in terms of the inferred linewidth ratio between the two lasers. Under the equal-contribution assumption, conventional two-laser heterodyne analysis reduces the inferred linewidth ratio to unity by construction~\cite{jiang2010nd,wu20160,hao2024stability}. \(\mathcal{M}a\)-function fitting instead treats this ratio as an observable encoded in the BLD, revealing departures from unity when the two lasers contribute unequally to the beatnote. For datasets independently benchmarked by TCH analysis, the \(\mathcal{M}a\)-function ratios agree well with the TCH-determined values~\cite{matei20171,chen2025laser,zhang2017ultrastable}. These comparisons establish the \(\mathcal{M}a\) function as a practical approach to individual linewidth characterization without requiring the equal-contribution assumption or assembling a metrologically equivalent TCH laser ensemble.
\par
In this work, we make essential steps towards resolving the long-standing question of the physical origin of the asymmetric BLDs first investigated by Ma more than sixteen years ago~\cite{jiang2010nd}. First, starting from the overlap time of photon-coherence wave trains from two lasers, we derive the \(\mathcal{M}a\) function, which reveals the physical origin of the observed asymmetry and is fundamentally distinct from conventional Gaussian and Lorentzian functions. Second, the \(\mathcal{M}a\) function enables direct extraction of the individual linewidths of two lasers from BLDs and accurately describes multiple experimental datasets~\cite{wu20160,zhang2017ultrastable,matei20171,hao2024stability}. TCH-type measurements with three cavity-stabilized lasers validate the \(\mathcal{M}a\) function and show that it resolves linewidth ratios that remain nearly degenerate in conventional TCH analysis. Third, we measure and fit the BLD between a cryogenic silicon-cavity-stabilized laser and a room-temperature ULE-cavity-stabilized laser. The method yields linewidths consistent with reported TCH results~\cite{chen2025laser}, demonstrating narrow-linewidth extraction without a metrologically equivalent reference ensemble. Overall, the \(\mathcal{M}a\) function opens a physically grounded, experimentally accessible, and resource-efficient avenue to accurate linewidth characterization of ultra-narrow-linewidth lasers, with broad relevance to optical clocks, gravitational-wave detection, high-resolution spectroscopy, and ultralow-phase-noise microwave generation. Ongoing work is extending this method to laser frequency-stability analysis.
\par

This work was funded by the National Natural Science Foundation of China (12503074), the Beijing Nova Program (No. 20240484696), the Wenzhou Major Science and Technology Innovation Key Project (No. ZG2023021), and the Quantum Science and Technology-National Science and Technology Major Project (No. 2021ZD0303200, 2021ZD0301903).

The authors thank L. S. Ma and H. Guan for their insightful discussions.

J. B. Chen proposed the concept of the beatnote-linewidth distribution function, namely the \(\mathcal{M}a\) function, and developed the corresponding linewidth-extraction method. D. Pan, J. M. Chen, and Y. C. Qi developed the theoretical model of the beat process. Y. C. Qi performed the numerical simulations. Y. Z. Pang contributed to the design and implementation of the software. C. Y. Wang and Y. Yao carried out the TCH experiment using two 1064 nm lasers and one 578 nm laser. Z. A. Chen and R. Q. Lei performed the heterodyne measurement between the 1397 nm cryogenic silicon-cavity laser and the room temperature ULE-cavity laser. J. M. Chen and Y. C. Qi wrote the manuscript and contributed equally to this work. All authors discussed the results and contributed to the revision of the manuscript.
\FloatBarrier
\bibliographystyle{apsrev4-2}
\bibliography{linewidth}
\end{document}